\documentclass[conference]{IEEEtran}

\usepackage{graphicx} 
\usepackage{tabularx}
\usepackage{float} 
\usepackage{amsmath}
\usepackage{booktabs}
\usepackage{longtable}
\usepackage{multirow}
\usepackage{algorithm}
\usepackage{algpseudocode}
\usepackage{cancel}
\usepackage{subcaption}
\usepackage{multicol}
\usepackage{url}
\usepackage{balance}
\usepackage{doi}
\usepackage{orcidlink}
\usepackage{wrapfig}
\usepackage[normalem]{ulem}

\usepackage{hyperref}
\hypersetup{
    colorlinks=true,
    linkcolor=black,
    filecolor=magenta,
    citecolor=black,
    urlcolor=blue,
}

\usepackage{colortbl, xcolor}
\usepackage[most]{tcolorbox}

\graphicspath{ {./figs/} }

\usepackage{listings}
\usepackage{tikz}
\usetikzlibrary{calc}
\usepackage{array}

\newcommand{\approach}{PerGent}
\newcommand{\baseline}{\textsc{OneShot}}
\newcommand{\woloop}{\textsc{OneShot+Res}}
\newcommand{\woresources}{\textsc{PerGentNoRes}}
\newcommand{\approachsc}{\textsc{\approach}}
\newcommand{\element}[1]{{\text{\fontsize{8.5}{8.5}\textsf{#1}}}}

\algrenewcommand\algorithmiccomment[1]{%
  \textcolor{green!50!black}{// #1}%
}

\definecolor{darkgreen}{RGB}{0,100,0}

\title{Agentic Persona Generation with Critique-Refinement: An Industrial Evaluation}

\author{
\IEEEauthorblockN{
Mohammad Hossein Amini\IEEEauthorrefmark{1}\IEEEauthorrefmark{2}\orcidlink{0009-0007-4312-7561},
David Dewar\IEEEauthorrefmark{2},
Shiva Nejati\IEEEauthorrefmark{1}\orcidlink{0000-0002-0281-8231},
Mehrdad Sabetzadeh\IEEEauthorrefmark{1}\orcidlink{0000-0002-4711-8319}
}
\IEEEauthorblockA{\IEEEauthorrefmark{1}
University of Ottawa, Canada\\
Email: \{mh.amini, snejati, m.sabetzadeh\}@uottawa.ca
}
\IEEEauthorblockA{\IEEEauthorrefmark{2}
Kinaxis, Canada\\
Email: ddewar@kinaxis.com
}
}

\begin{document}

\maketitle

\pagestyle{empty}

\begin{abstract}
Personas are widely used in software engineering to support requirements elicitation, design, and validation, but their manual creation is costly, time-consuming, and hard to scale. Recent LLM-based approaches automate persona generation from textual data; however, they typically rely on single-shot generation and subjective evaluations, limiting practical reliability. We present \approach, an industry-grade method for persona generation built around an iterative critique-refinement loop. Specifically, \approach\ uses a generator and a critic LLM agent, coordinated by an orchestrator, to iteratively refine personas using external resources such as interviews, surveys, and job postings through a critique-refinement loop with a user-defined maximum number of rounds. We deploy and evaluate \approach\ in an industrial setting at Kinaxis, comparing it with three baselines, including  one-shot methods. In an expert in-situ evaluation, \approach\ achieved the highest expert approval rate (96.9\%), exceeding all baselines. We further compare \approach-generated personas with best-practice personas manually created by domain experts prior to the adoption of LLMs. Compared to baselines, \approach\ reproduces a larger proportion of expert content while also contributing substantial new content beyond the pre-LLM personas. We conclude with lessons learned from deploying and evaluating \approach\ at Kinaxis.
\end{abstract}


\section{Introduction}
\label{sec:intro}

Personas play an important role in software engineering by representing user archetypes and guiding requirements elicitation, design decisions, and validation~\cite{Salminen_2025_PersonaCraft, salminen2024deus, Chetan_2025_Who_uses_personas, Chetan_2023_PersonaGen, Cooper1999Inmates}.
Persona templates typically include technical, demographic, and behavioural dimensions and are tailored to the domain and usage context to keep requirements grounded in real-world needs~\cite{Chetan_2025_Who_uses_personas, Chetan_2023_PersonaGen, Salminen_2025_PersonaCraft}.

Persona generation is difficult in practice, as it requires extensive synthesis of heterogeneous user-research data~\cite{Salminen_2025_PersonaCraft}. To create a persona manually, experts must examine diverse user-research artifacts, such as interview transcripts that inform persona attributes, survey data that validate patterns across roles and responsibilities, and internal documentation that provides domain and organizational context. This process is tedious and costly; under tight deadlines, practitioners often produce personas that fall short, for example, by being incomplete, overly generic, or insufficiently actionable~\cite{Salminen_2025_PersonaCraft, Shin_2024_Humain_AI_Workflows}.

Before large language models (LLMs), researchers attempted to automate persona generation using machine learning and data analytics applied to quantitative platform metrics such as view counts, engagement frequency, audience shares, and inferred demographic attributes~\cite{An_2018_Imaginary_People, Jansen_2020_Personas_and_analytics}.
Since this strategy depends on statistical patterning and clustering over structured numeric features, it has difficulty incorporating unstructured qualitative text (e.g., interview narratives, open-ended survey responses, and observational notes), which is essential for producing practically useful personas.

LLMs mitigate the above limitation by enabling the use of rich textual data sources for automated persona generation~\cite{jung2018apg,schuller2024llm,salminen2024deus,Salminen_2025_PersonaCraft,Chetan_2023_PersonaGen}. However, existing LLM-based persona-generation methods still have two key  shortcomings.
\emph{First}, they generate personas in a single LLM call (single-shot) without iterative refinement, so experts must post-edit the results~\cite{Shin_2024_Humain_AI_Workflows}. By comparison, recent agentic LLM approaches in software engineering iteratively improve artifacts through a critique-refinement loop, where one LLM agent evaluates a draft and identifies problems while another attempts to address them~\cite{Pradel_2025_RepairAgent,Batole_2025_Agent_IssueLocalization,Cheng_2025_TSEMultiAgentSmartContract}. Empirical studies in requirements analysis and code synthesis show these loops reduce hallucinations and improve factual grounding~\cite{Roy_2024_Root_Cause,Parham_2025_Critique,AyoughiICSESEIP2026}. To our knowledge, LLM-based persona generation has not yet adopted this critique-refinement loop and remains single-shot.

\emph{Second}, existing evaluations of LLM-generated personas are largely perception-based, focusing on subjective judgments of plausibility or overall persona quality~\cite{salminen2024deus,schuller2024llm}. By contrast, the \emph{practical usefulness} of LLM-generated personas in real organizational contexts has not been systematically evaluated. In particular, there is limited practice-based evidence derived from observing how domain experts employ these personas and from controlled comparisons between LLM-generated personas and best-practice personas developed manually within organizations. Without such evidence, it remains unclear whether LLM-generated personas can reliably support expert decision-making or be suitable for professional use.

In this paper, we introduce \emph{\approach}, an agentic method for persona generation built around a critique–refinement loop. \approach\ consists of two LLM agents, a persona generator and a persona critic, coordinated by an orchestrator. Given a target persona description and relevant external resources, e.g.,~user interviews and surveys, the generator produces an initial persona and iteratively refines it in response to the critic’s feedback. The critic assesses each draft against criteria such as completeness, coherence, and relevance, flagging potential gaps or inconsistencies. The critique-refinement loop runs for a user-defined number of rounds or stops earlier when the critic indicates that no major issues remain.

Our evaluation has three complementary components:

The \emph{first} component evaluates \approach\ through an in-situ industrial deployment at Kinaxis. In this deployment, \approach\ alongside three baseline techniques -- two of which are one-shot methods from the literature~\cite{Salminen_2025_PersonaCraft, Chetan_2023_PersonaGen} -- were instantiated as four tools within a shared operational workflow. Four Kinaxis domain experts used these tools in their daily work for approximately one month. We examine interaction logs that record expert approvals and revisions to persona content, assessing how well the persona content aligns with expert expectations in industrial applications. The results of our in-situ deployment indicate that \approach\ generates personas that closely match expert expectations and require minimal post-editing: experts approved 96.9\% of the generated content unchanged, made no additions or removals, and modified only 3.1\% of the content, primarily to correct outdated details. In comparison, baseline LLM-based persona generators required more expert intervention, with unchanged-content approval rates ranging from 75.8\% to 93.9\%, and with some baselines requiring non-trivial additions or removals. Across revisions, most changes replaced outdated details with current information and were mainly needed when no external sources were provided, underscoring the importance of grounding persona generation in up-to-date sources.

Our \emph{second} evaluation component compares personas generated by \approach\ and the three baselines against a best-practice reference set of personas authored by Kinaxis domain experts prior to the adoption of LLMs. We investigate two central questions: (i)~to what extent can each automated method \emph{reproduce} this expert-authored content? and (ii)~to what extent can it contribute additional (novel) insights beyond the expert-authored content?
Accordingly, we report two metrics: \emph{preservation} (expert-material recovery) and \emph{distinctness} (novel content beyond the reference). To assess these metrics reliably, we evaluated seven LLMs (GPT-4o, o4-mini, GPT-5.2, Gemini-2, Gemini-3, Claude-4, and Grok-3). For every combination of a method, an LLM, and a job role, we generated five personas to reduce randomness (1,400 total persona generations). The results show that \approach\ consistently outperforms the baselines on both metrics, achieving up to an average of 9.5\% higher preservation and up to an average of 14.2\% higher distinctness than the baselines.

Our \emph{third} evaluation component measures the computational cost of \approach\ and the baselines in terms of LLM calls per persona and input/output tokens per call. Across the seven LLMs considered, \approach\ typically uses multiple calls to support its critique-refinement process (4.8 calls per persona on average, with roughly 11K input tokens and 1.7K output tokens per call), while the one-shot baselines generate personas in a single call with lower token usage. 

We conclude the paper by presenting the key lessons derived from our empirical findings.

\textbf{Contributions.} Our main contributions are as follows:

\textbf{(1) Agentic persona generation.} We present \approach, an agentic persona generation method structured around iterative critique and refinement. \approach\ is developed as an orchestrator-mediated multi-agent system using the Microsoft AutoGen framework~\cite{Wu_2023_Autogen}.

\textbf{(2) In-situ expert evaluation.} We evaluate \approach\ and three baseline methods by deploying them in an industrial setting and studying their use by domain experts over a one-month period, using both quantitative and qualitative analyses. We show that personas generated by \approach\ require minimal expert intervention, achieving a 96.9\% item approval rate and outperforming the baselines.

\textbf{(3) Comparison with pre-LLM personas.} 
We present a systematic evaluation comparing LLM-generated personas with personas created by domain experts prior to the adoption of LLMs. Our results show that \approach\ significantly outperforms baseline methods in reproducing expert-authored content from pre-LLM personas while also introducing useful, non-redundant additions.

\section{Industrial Context}
\label{sec:context}
Kinaxis provides supply-chain planning software for global enterprises  across a wide range of industries. Because these industries involve complex workflows and highly specialized roles, effective requirements elicitation and structured requirements definition require a thorough understanding of users' tasks, the challenges they encounter, and the tools they rely on in their daily work. To support requirements-related activities, including requirements identification, prioritization, and the engineering of user experience (UX) requirements, Kinaxis employs \emph{role-based personas}~\cite{Chetan_2025_Who_uses_personas}, such as demand planners and inventory managers. These personas function as representations of real users, helping ensure that defined requirements and the resulting system capabilities remain consistently aligned with user needs, \hbox{pain points, and operational contexts.}

\begin{figure}[t]
    \centering
    \includegraphics[width=0.85\columnwidth]{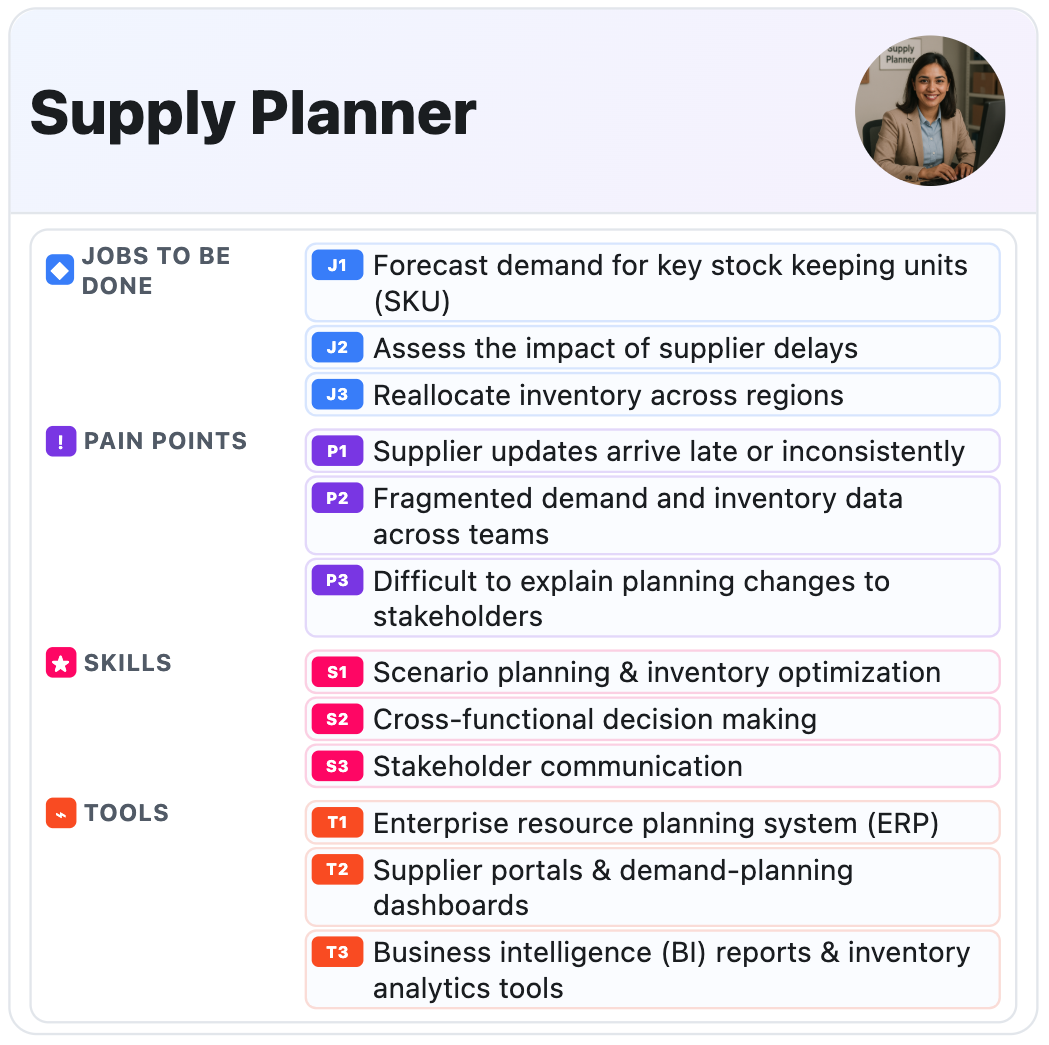}
	\caption{Example of a supply-planner persona; each section may include up to ten items (only three shown). Items are labelled with identifiers to facilitate referencing.}
    \label{fig:persona_sample}
\vspace*{-1.5em}
\end{figure}

At Kinaxis, a persona is structured into four sections: \emph{jobs to be done}, describing what a person is responsible for and aims to achieve; \emph{pain points}, outlining the challenges they face; \emph{required skills}, capturing the capabilities needed for their role; and \emph{tools used}, indicating the resources and systems they rely on. Although personas are often associated with user-interface and interaction design~\cite{Cooper1999Inmates}, they have increasingly been adopted to support requirements engineering (RE) activities, including requirements elicitation and analysis, requirements specification, and requirements validation~\cite{karolita2023use}.
The four-section structure used at Kinaxis is consistent with recurring elements of personas in the RE literature, including users' goals and tasks, concerns or pain points, skills and domain knowledge, and interactions with tools and technologies~\cite{karolita2023use,karolita2023whats}.

In RE, such information helps analysts reason about what users need to accomplish, where current workflows break down, what capabilities users bring to their work, and which technical and organizational constraints shape feasible requirements. We note that persona representations are not fixed: their structure and content may vary depending on the domain, organizational practices, and intended use~\cite{karolita2023whats, Chetan_2025_Who_uses_personas}. Our goal here is therefore not to advance the Kinaxis structure as a general persona template, as other organizations may adopt different templates depending on their context. Instead, our aim is to examine automated persona generation in an established industrial workflow.

We illustrate through an example how personas expressed in our industry partner's template support RE tasks. Figure~\ref{fig:persona_sample} shows a sample supply-planner persona. The \emph{jobs to be done} items indicate that planners must assess the impact of supplier delays (J2) and reallocate inventory across regions (J3). This suggests a need for requirements concerning the comparison of alternative supply-allocation scenarios. The \emph{pain points} further refine the considerations to be accounted for when defining such requirements: supplier updates may arrive late or inconsistently (P1), relevant demand and inventory data may be fragmented across teams (P2), and planning changes may be difficult to explain to stakeholders (P3). These items suggest that any functionality specified by the requirements should not introduce an isolated planning mechanism, but should instead consolidate supplier-status, demand, and inventory information and make changes from previous plans explicit. 
The \emph{skills} and \emph{tools} sections further ground requirements in the user's expertise and existing work environment. In our example, the \emph{skills} items indicate that the system should support scenario planning, cross-functional decision making, and stakeholder communication (S1--S3), for instance by explaining the assumptions and trade-offs behind alternative plans. Finally, the \emph{tools} items indicate that the user need should be considered in relation to the existing ERP system, supplier portals and demand-planning dashboards, and BI reports and inventory-analytics tools (T1--T3). 

The persona-based analysis illustrated above makes requirements elicitation and specification more concrete and context-aware. For example, based on this analysis, stakeholders could converge on the need for ``the system to support scenario-based supply reallocation by integrating supplier-status, demand, and inventory data from existing tools, presenting the assumptions and trade-offs of alternative plans, and enabling planners to communicate selected changes.'' This statement of need can then be further refined through subsequent RE activities, including additional elicitation and negotiation, before being translated into fine-grained requirements.%

Manually creating and maintaining personas requires significant time and coordination, and can become increasingly complex as customer bases and use cases grow. In 2022, Kinaxis undertook a strategic initiative to formalize and scale its persona development practice: Five senior product managers led the development of ten personas\footnote{In our paper, we refer to these personas as \emph{pre-LLM personas} and compare the personas generated using \approach\ against these personas in Section~\ref{sec:empirical_eval}.}. The work spanned nearly a year and began with an intensive initial phase of scoping, drafting, cross-team reviews, refinement, and final approval. Each persona required synthesizing heterogeneous inputs (e.g., interviews, surveys, internal playbooks, and job postings), aligning terminology across roles and regions, and iteratively clarifying overlap and boundaries among personas.

Following this initial phase, further expansion and ongoing updates (e.g., to reflect new industries, regions, and release cycles) would have required a comparable level of expert effort and sustained organizational commitment. Building on this past initiative, Kinaxis identified an opportunity to streamline persona generation and maintenance at scale, leading to the exploration of LLMs to support this process. 
This paper aims to design and systematically evaluate an agentic method for automatically generating reliable personas by drawing on a broad range of context-relevant resources from Kinaxis.

\section{Agentic Persona Generation}
\label{sec:methodology}
In this section, we introduce \approach, our agentic persona generation method. We begin in Section~\ref{subsec:design} by describing the agentic design adopted in our work, which is derived from Microsoft’s AutoGen framework~\cite{Wu_2023_Autogen}. We then detail, in Section~\ref{subsec:pergent}, how \approach\ instantiates this design.

\subsection{Orchestrated Multi-Agentic Design}
\label{subsec:design}
\begin{figure}[t]
   \centering
      \includegraphics[width=0.8\linewidth]{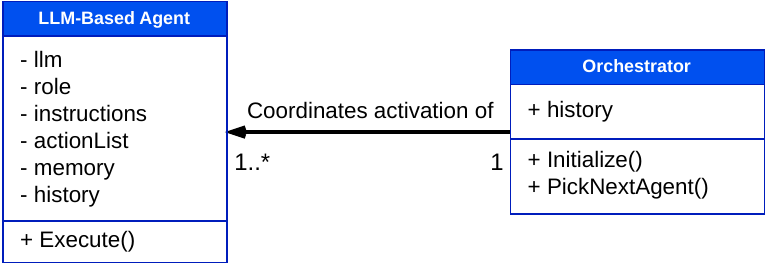}
   \caption{Structure of our orchestrated multi-agentic design based on  Microsoft’s AutoGen framework~\cite{Wu_2023_Autogen}.}
   \label{fig:architecture}
\end{figure}

Figure~\ref{fig:architecture} presents a class diagram characterizing the structure of our agentic design, inspired by the principles of Microsoft's AutoGen framework~\cite{Wu_2023_Autogen}. While AutoGen supports different agentic designs with diverse orchestration strategies and interaction patterns among agents and the orchestrator, we adopt a \emph{centralized, turn-based orchestration strategy} that is sufficient for implementing the critique-refinement loop of \approach. Our agentic design consists of a single orchestrator and a set of LLM-based agents.
The orchestrator determines agent turn order using a round-robin policy and records agent outputs using its \element{history} attribute, while each agent independently selects its actions. The orchestrator terminates execution when an agent signals completion or a user-defined maximum number of round-robin cycles is reached. Our design aligns with a \emph{Level~3} goal-agentic configuration in the taxonomy of Hassan et al.~\cite{hassan2025agenticSE}, where agents act autonomously within an orchestrator-mediated setup.

The attributes of the \element{LLM-Based Agent} class (Figure~\ref{fig:architecture}) collectively define how an  agent is set up and how it decides which actions to perform.  While agents can perform a wide range of actions defined in their \element{actionList} attribute, potentially involving external tools, each agent offers a single external interface, \element{Execute()}. When invoked, this method, whose behaviour is outlined in Algorithm~\ref{alg:execute}, determines  which action to perform. Specifically, the \element{Execute()} method first synchronizes the agent’s history with that of the orchestrator, ensuring that the agent has access to outputs produced during previous invocations, whether by itself or by other agents (line~2). It then constructs a prompt using the agent’s attributes, namely \element{role}, \element{instructions}, \element{actionList}, \element{memory}, and \element{history}, and submits this prompt to the underlying LLM (lines~3–4). Guided by the agent’s role, instructions, memory, and history, the LLM selects and executes an appropriate action from the \element{actionList}. The resulting output is appended to the agent’s \element{history}, after which the orchestrator’s \element{history} is updated to reflect this result, ensuring that subsequent agent invocations can access the output generated in this execution (lines~5–6).

\begin{algorithm}[t]
\caption{\textsc{Execute}() method of the \element{LLM-Based Agent} class}
\label{alg:execute}
{\scriptsize
\begin{algorithmic}[1]
\Function{Execute()}{}
  \State history $\gets$ orchestrator.history
  \State prompt $\gets$ \Call{BuildPrompt}{role, instructions, actionList, memory, history}
  \State output $\gets$ \Call{CallLLM}{llm, prompt}
    \State history $\gets$ history $\cup \{ \text{output} \}$
  \State orchestrator.history $\gets$ history
\EndFunction
\end{algorithmic}}
\end{algorithm}

\subsection{\approach}
\label{subsec:pergent}
\begin{figure}[t]
   \centering
   \includegraphics[width=\columnwidth]{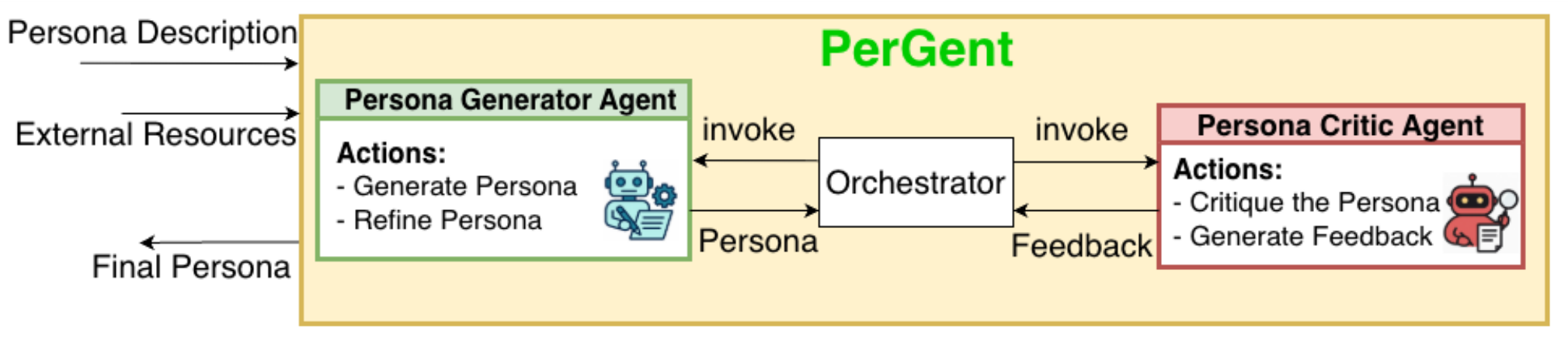}
   \caption{Overview of our agentic persona generator (\approach).}
   \label{fig:approach}
\end{figure}

Figure~\ref{fig:approach} provides an overview of \approach, which takes as input a target persona description and external resources collected from practitioners (e.g., surveys and interviews). It then generates a persona through a critique-refinement loop consisting of an orchestrator, a generator, and a critic agent. The orchestrator of \approach\ operates according to the process described in Section~\ref{subsec:design}. The instantiations of the attributes of \approach's generator and critic agents are provided in Table~\ref{table:prompts}. Both agents follow the implementation of the \element{Execute()} method, as described in Algorithm~\ref{alg:execute}. Below, we outline the two agents. Further details about the implementation of \approach\ are provided in Section~\ref{subsec:impl}. The prompts constructed by each agent's \element{Execute()} method are provided online~\cite{Github}.

\begin{table}[t]
\centering
\scriptsize
\caption{Instantiating the \element{LLM-Based Agent} in Fig.~\ref{fig:architecture} for the Generator and Critic agents of \approach.}
\label{table:prompts}
\scalebox{0.75}{%
\begin{tabular}{p{2cm} | p{4.2cm} | p{4.0cm}}
\toprule
\textbf{Agent Attribute} & \textbf{Generator Agent} & \textbf{Critic Agent} \\
\midrule
\element{role} & Persona generator and refiner & Persona critic \\
\midrule
\element{instructions} & Instructions on generating and refining personas; syntactic definition of persona (required sections, number of items, etc.) & Instructions on evaluating personas; syntactic definition of persona (required sections, number of items, etc.) \\
\midrule
\element{actionList} & (1)~Generate a persona from scratch; (2)~Refine a persona & (1)~Critique a persona; (2)~Generate feedback \\
\midrule
\element{memory} & \multicolumn{2}{c}{External resources and target persona description} \\
\midrule
\element{history} & \multicolumn{2}{c}{List of previously generated personas and feedback} \\
\midrule
\element{output} & A persona & Feedback or approval declaration \\
\bottomrule
\end{tabular}}
\vspace*{-.4cm}
\end{table}

The \textbf{generator agent} creates and refines personas. Its \element{instructions} define the required persona format and quality guidelines. The agent's \element{memory} stores the inputs to \approach, namely the target persona description and external resources. The agent supports two actions: generating an initial persona when \element{history} is empty, and refining an existing persona based on critic feedback and prior drafts recorded in \element{history}. The \textbf{critic agent} evaluates the generated personas. Its \element{instructions} define the persona format and evaluation criteria. Like the generator agent, its \element{memory} stores \approach’s inputs. In each execution, the critic assesses the latest persona against these inputs and either provides feedback  or approves the persona.

\section{Empirical Evaluation}
\label{sec:empirical_eval}

Our evaluation addresses three research questions (RQs):

\textbf{RQ1 (Expert validation).} \emph{How much expert revision do personas generated by \approach\ require before approval?} We evaluate how domain experts at Kinaxis review and refine personas generated by \approach\ and by baseline methods using a one-month log of their interactions. The log records every edit, removal, addition, and approval applied to the generated personas. We analyze this review process both \emph{quantitatively} and \emph{qualitatively}. Quantitatively, we measure approval and modification frequencies and assess how closely personas from \approach\ and the baselines align with expert expectations. Qualitatively, we examine the types of changes and the motivations behind them.

\textbf{RQ2 (Validation against pre-LLM personas).} \emph{How do \approach-generated personas compare with pre-LLM, expert-built personas?} We evaluate personas generated by \approach\ and the baselines against pre-LLM personas developed by domain experts, whose development was discussed in Section~\ref{sec:context}. Our comparison focuses on (i)~the extent to which the LLM-generated personas reproduce (i.e., \emph{preserve}) expert-generated content present in the pre-LLM personas, and (ii)~the extent to which they contribute novel information (i.e., \emph{distinct} from what is already known) relative to the pre-LLM personas. The metrics we introduce to quantify (i) and (ii) are described in Section~\ref{subsec:metrics}.

\textbf{RQ3 (Cost).} \emph{
What is the cost of \approach\ in terms of LLM calls and tokens per call?
} We compare \approach\ and the baselines on persona-generation cost using the number of LLM calls and tokens required to generate personas. 
RQ3 complements the quality-focused analyses in RQ1 and RQ2 by quantifying the deployment cost of each persona-generation strategy. This cost is important from a practical standpoint because personas may need to be generated and maintained across many roles, products, and evolving resources. RQ3 provides the empirical basis for the trade-off we discuss in Lesson~1 of Section~\ref{sec:lessons}, namely, whether the quality gains from critique-refinement and external resources justify the additional per-persona LLM-call and token overhead.


\subsection{Persona Generation Workflow and Baselines}
\label{sec:baseline}
\begin{figure}[t]
   \centering
   \includegraphics[width=\linewidth]{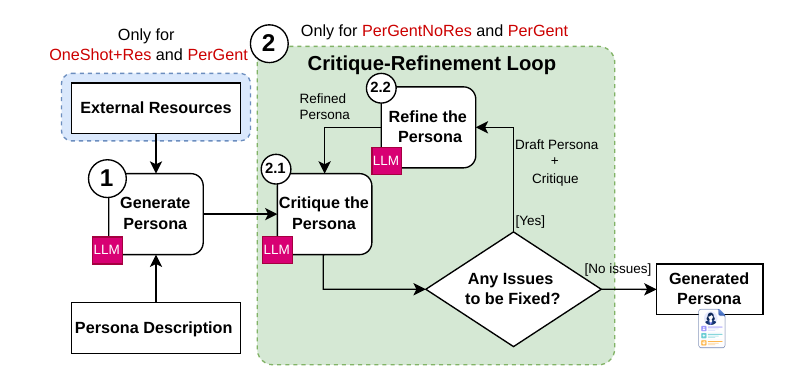}
\caption{Workflow of \approachsc\ and the baselines. \woresources\ uses both Steps~1~and~2, but without external resources; \baseline\ and \woloop\ use only Step~1, with external resources used only by \woloop.}
\label{fig:tool}
   \vspace*{-1em}
\end{figure}

Figure~\ref{fig:tool} presents our persona generation workflow. \approach\ follows the agentic design described in Section~\ref{sec:methodology}, with its operational steps shown in this figure. Given a persona description and a set of external resources, the generator agent first produces a persona draft (Step~1 in Figure~\ref{fig:tool}). The critic agent then evaluates the draft (Step~2.1 in Figure~\ref{fig:tool}). If the critic identifies issues, the generator refines the draft based on the critic's feedback (Step~2.2 in Figure~\ref{fig:tool}). This critique-refinement loop continues until the critic reports no remaining issues, or until a predefined iteration cap is reached, at which point the final persona is returned.
We use \textsc{\approach} to refer to the concrete implementation of \approach\ in our evaluation.

We develop three baselines -- \woresources, \baseline, and \woloop\ -- as variations of the workflow in Figure~\ref{fig:tool}. \baseline\ and \woloop\ represent single-shot LLM-based persona generation approaches from the literature~\cite{Chetan_2023_PersonaGen, Salminen_2025_PersonaCraft}. These two approaches perform persona generation only (i.e., Step~1) without critique-refinement. \baseline\ uses only the target persona description as input, whereas \woloop\ uses both the target persona description and external resources. \woresources\ is a variant of \approachsc\ that keeps the critique-refinement loop but removes external resources, using only the target persona description as input.
 
  \baseline\ and \woloop\ evaluate the impact of the critique-refinement loop, whereas \woresources\ isolates the effect of external resources on generated personas. \woresources\ is motivated by the fact that curating external resources, such as interviews and surveys, can be costly and may not be feasible in all contexts.  Through this baseline, we evaluate the contribution of external resources to persona quality within an agentic architecture.

\subsection{Implementation} 
\label{subsec:impl}
\approachsc\ is implemented in Python~3.13 using the Microsoft AutoGen~0.7.5 framework~\cite{Wu_2023_Autogen}. 
To ensure a fair comparison with \approachsc, we implemented our three baselines as follows: \woresources\ reuses the same AutoGen-based  implementation of \approachsc\ but with the external resources removed from its inputs. 
In our experiments, we use the same LLM for both agents (generator and critic) in \woresources\ and \approachsc.
\baseline\ and \woloop, as they only execute Step 1 in Figure~\ref{fig:tool}, follow a simpler implementation without AutoGen and invoke their underlying LLM  directly through API calls. To ensure consistency, for \baseline\ and \woloop, we used the same persona-generation instructions, section definitions, output-format constraints, and examples as the generator agent in \approachsc. The prompt outlines and implementations of \approachsc\ and all three baselines are provided in our online repository~\cite{Github}.

To support our RQ1 evaluation and facilitate company-wide adoption at Kinaxis, we developed a production-grade user interface implemented in Streamlit~1.54. The interface was iteratively refined over the course of a year in close collaboration with the Kinaxis UX team to ensure alignment with internal work practices. The same interface was used for \approachsc\ as well as for the three baselines, 
yielding four tools:  one for \approach\ and three for the baselines.
Each tool enables an expert to automatically generate a persona and subsequently review and refine it through edit, addition, removal, and approval actions, all of which are recorded using \hbox{Python's built-in logging module.}

\subsection{External Resources}
In our evaluation, we consider ten job roles covering the primary supply-chain functions, from planners (e.g., supply and demand planners) to administrators and managers. These resources fall into one of the following three types:
(1) \emph{One-on-one interviews}: transcripts of interviews between a Kinaxis requirements analyst (domain expert) and a practitioner in the target role. The analyst asks about tasks, pain points, required skills, and commonly used tools. Interviews averaged 58.3 minutes, with 89 analyst turns and 88 practitioner turns. Analysts spoke an average of 1,794 words and interviewees 7,353 words.
(2) \emph{Surveys}: structured questionnaires completed by practitioners, capturing self-reported tasks, challenges, skills, and tools; each survey contained 23 structured questions (the same instrument for all roles).
(3) \emph{Job postings}: role-specific postings curated from the supply-chain job market, reflecting industry-standard responsibilities, required skills, and commonly used tools.
Across the ten roles, Kinaxis provided a total of 25 interview transcripts, 28 surveys, and 47 job postings.
We note that, when generating a persona, \approachsc\ and \woloop\ receive only resources relevant to that persona; for example, an interview with a supply-chain manager is never provided when generating a demand-planner persona.

\subsection{Pre-LLM Personas}
\label{subset:prellm}
As described in Section~\ref{sec:context}, five senior product managers at Kinaxis manually developed ten personas in 2022, without using LLMs. We refer to this set of personas as \emph{pre-LLM} personas and use them to answer RQ2. The pre-LLM personas all consist of the same four sections as the personas generated by \approachsc\ and the baselines, i.e., \textit{jobs to be done}, \textit{pain points}, \textit{skills}, and \textit{tools}.
In total, the ten pre-LLM personas comprise 382 items (38.2 items per persona on average), with a balanced distribution across the four sections. The average and standard deviation of words \hbox{per item are 13.96 and 7.43, respectively.}

The pre-LLM personas reflect expert knowledge at a specific point in time and constitute a reference set rooted in real industrial practice, as they were manually created through deliberate effort by individuals with deep domain expertise. They therefore provide a meaningful basis for assessing the extent to which LLM-based approaches reproduce and expand expert knowledge. However, because the personas were developed in 2022, some items have become invalid due to  changes in tools and workflows. For example, certain listed tools are no longer in use, which could affect the reliability of the RQ2 evaluation. To address this, two domain experts reviewed the personas and removed outdated or irrelevant items (less than 5\% of all items) prior to our evaluation. Apart from these removals, all personas and their overall structure -- including all sections -- remained unchanged.

\subsection{Metrics}
\label{subsec:metrics}
\underline{\textit{Metrics for RQ1.}} We inspect expert interactions  through the logs collected from the deployed tools for \approach\ and the baselines. 
For each generated persona, we compute the percentage of items that are (i)~edited, (ii)~removed, (iii)~added, and (iv)~approved without modification. Percentages are calculated relative to the total number of items in the finalized persona.
These measures capture how closely LLM-generated personas align with expert expectations in real industrial use and directly reflect the post-generation human effort required.

\vspace*{.3em}

\underline{\textit{Metrics for RQ2.}}
We define two metrics to compare personas generated by \approachsc\ or the baselines \emph{against pre-LLM personas}.
Our goal here is to measure how well an LLM-generated persona \emph{preserves} the original content of the pre-LLM persona and how much \emph{distinct} content it adds beyond the pre-LLM persona. 
To our knowledge, no prior work has directly compared the content of two personas. Existing evaluation approaches~\cite{Chetan_2023_PersonaGen, Salminen_2025_PersonaCraft, Jansen_2020_Personas_and_analytics, salminen2024deus} typically assess individual personas in isolation (as in our RQ1), rather than measuring how well one persona covers or expands on another. As a result, no established metrics are available for comparing an LLM-generated persona against its corresponding pre-LLM persona; this gap motivates the metrics we introduce here.

Recall that, in our setting, each persona is a set of items, each of which captures concepts from one or more of four categories: jobs to be done, pain points, skills, and tools. 
From now on, we refer to the concepts related to these four categories as \emph{persona-related concepts}.
To formally define metrics for RQ2, we first describe how the persona-related concepts in an individual item of one persona can either be \textit{fully present}, \textit{partially present} or \textit{not present} in the entire content of another persona.
We then define two metrics, namely \emph{distinctness} and \emph{preservation}, based on the item-persona relationships to compare an LLM-generated persona with a pre-LLM persona.

\textbf{Item-Persona Relationship.}
Let $P$ denote the set of items in one persona and $Q$ denote the set of items in another persona. 
We characterize how a single item $p \in P$ relates to the \textit{other persona as a whole} ($Q$) using three levels: \textit{fully present}, \textit{partially present}, and \textit{not present}.  Table~\ref{table:relations} defines these three levels and provides examples for each.

\begin{table}[t]
\begin{center}
{\footnotesize
\caption{Possible relationships between an item $p$ of a persona $P$ and another persona $Q$, defined by the extent to which the persona-related concepts (jobs to be done, pain points, skills, and tools) mentioned in $p$ are presented in $Q$.}
\label{table:relations}
\scalebox{.9}{
\begin{tabular}{|p{1.85cm}|p{2.5cm}|p{4cm}|}
\hline
\textbf{Relationship} & \textbf{Definition} & \textbf{Example} \\ \hline

Fully Present &
All persona-related concepts mentioned in $p$ appear in $Q$, possibly across multiple items or with different wording. &
Item $p$: ``\textcolor{darkgreen}{Inventory levels are often too low or too high.}''  
Persona $Q$ contains: ``\textcolor{darkgreen}{Products run out unexpectedly}'' and 
``\textcolor{darkgreen}{Excess items remain in storage.}''\\ \hline

Partially Present &
Only some of the persona-related concepts mentioned in $p$ appear in $Q$; the rest are missing. &
Item $p$: 
``\textcolor{blue}{Inventory levels are often too low or too high.}''
Persona $Q$ contains: ``\textcolor{blue}{Products run out unexpectedly.}'' \\ \hline

Not Present &
None of the persona-related concepts mentioned in $p$ appear anywhere in $Q$. &
Item $p$: 
``\textcolor{magenta}{Inventory levels are often too low or too high.}''  
Persona $Q$ contains no item related to inventory levels. \\ \hline

\end{tabular}
}
}
\end{center}
\vspace*{-.7cm}
\end{table}

We define the \emph{distinctness} and \emph{preservation} metrics between an LLM-generated persona, denoted by $P$, and a pre-LLM persona, denoted by $Q$, as follows:

\textbf{Distinctness.}
An item $p \in P$ is counted as \emph{distinct} if it is either \textit{not present} or \textit{partially present} in $Q$.
\emph{Distinctness} is the percentage of \emph{distinct} items in the LLM-generated persona.
For example, a distinctness of 30\% means that 30\% of the items in the LLM-generated persona contain concepts not presented in the corresponding pre-LLM persona, while the remaining 70\% add nothing new compared to the pre-LLM persona.

\textbf{Preservation.}
To provide a more nuanced view of preservation, we define two variants of this metric: (1)~\emph{full preservation} and (2)~\emph{partial preservation}.
\emph{Full preservation} is the percentage of items in the pre-LLM persona $Q$ that are \textit{fully present} in the LLM-generated persona $P$, i.e., all persona-related concepts are retained.
\emph{Partial preservation} is the percentage of items in $Q$ that are \textit{fully or partially present} in $P$. This captures cases where some original concepts remain despite incomplete coverage, potentially cueing analysts to missing information. For example, if full preservation is 40\%, then 40\% of items are completely retained; if partial preservation is 70\%, an additional 30\% have partial coverage.

\pagebreak

\underline{\textit{Metrics for RQ3.}} We measure computational cost using two metrics: (i)~the number of LLM calls to generate a persona and (ii)~the number of input and output tokens per call.

\subsection{RQ1 Experiments and Results} 
\label{subsec:rq1}

\begin{figure*}[t]
   \centering
   \includegraphics[width=.992\linewidth]{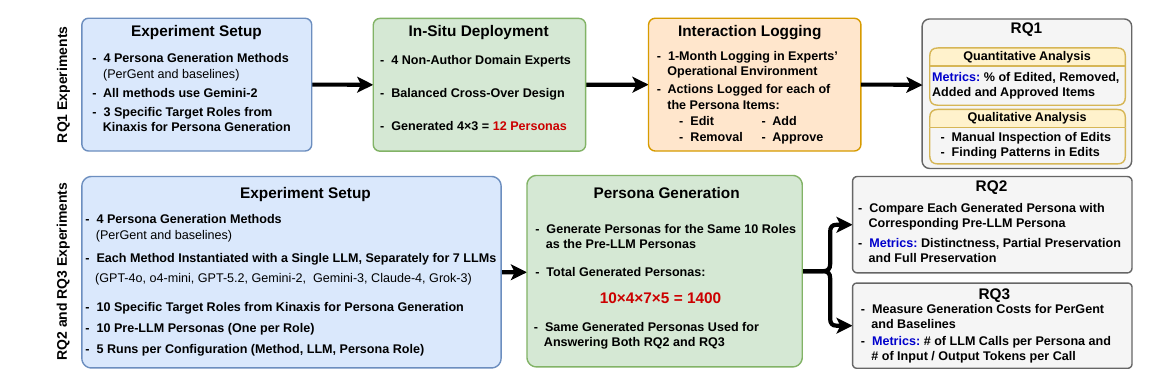}
   \caption{Experiment workflows for our research questions (RQ1, RQ2 and RQ3)}
   \label{fig:exp_workflow}
   \vspace*{-0.7em}
\end{figure*}

\subsubsection{RQ1 Experiments}  
Figure~\ref{fig:exp_workflow} (top) shows the workflow of the RQ1 experiments.
To collect expert feedback, we deployed \approachsc\ and the three baselines as independent tools (see Section~\ref{subsec:impl}) at Kinaxis.  For the purpose of answering RQ1, we configure all four tools to use Gemini-2, Kinaxis's primary enterprise LLM.  Each persona generated with these tools follows the structure illustrated in Figure~\ref{fig:persona_sample}: a title, an avatar, and four sections containing multiple items as specified in the figure. Each tool was instrumented to automatically log all user interactions -- capturing every persona generated, all modifications (edits, additions and removals) and all approvals without modifications.
To keep the review and refinement workload manageable for the participating experts, persona generation in RQ1 was limited to three common supply-chain roles: \textit{supply planner}, \textit{demand planner}, and \textit{capacity planner}. 

\begin{table}[t]
\centering
\caption{Cross-over assignment of experts  to tools across persona roles; each cell shows the tool an expert used to generate and refine the persona.}
\label{tab:crossover}
\begin{tabular}{c|ccc}
\toprule
 & \textbf{Role 1} & \textbf{Role 2} & \textbf{Role 3} \\
\midrule
\textbf{Expert 1} & Tool 1 & Tool 2 & Tool 3 \\
\textbf{Expert 2} & Tool 2 & Tool 3 & Tool 4 \\
\textbf{Expert 3} & Tool 3 & Tool 4 & Tool 1 \\
\textbf{Expert 4} & Tool 4 & Tool 1 & Tool 2 \\
\bottomrule
\end{tabular}
\vspace*{-.5cm}
\end{table}

To examine how domain experts interacted with \approachsc\ and the baselines, we provided the implemented tools to four Kinaxis experts: two senior UX specialists and two senior product managers. The experts had extensive professional experience working with clients in the three roles of our study (i.e., supply planner, demand planner, and capacity planner). All four experts further had considerable prior experience with requirements elicitation and persona development.

We deployed the four tools in the experts' operational environment, enabling them to generate and refine personas for the three aforementioned roles using fixed inputs (persona descriptions and, where applicable, external resources).
To mitigate confounding effects, we adopted a cross-over design, summarized in Table~\ref{tab:crossover}. Under this design, each of the four experts generated and refined one persona for each of the three roles, using a different tool for each role. Experts were blinded to the underlying persona-generation method associated with each tool because all tools shared an identical UI (see Section~\ref{subsec:impl}).
As shown in Table~\ref{tab:crossover}, the assignment was balanced such that (i) each tool was used by three different experts, (ii) each expert interacted with three different tools, and (iii) for each role, four personas were produced -- one per tool and one per expert. 

To answer RQ1, we analyzed logs from a one-month period during which all four experts were expected to engage with the tools as part of their day-to-day tasks. After this period, we collected the logs generated by each tool to characterize how each expert interacted with and improved the personas. The logs indicated that all experts had completed their tasks, yielding a total of twelve ($4 \times 3$) personas.

\begin{table}[t]
\centering
\caption{Summary statistics for expert-validated personas.}
\label{tab:rq1-persona-summary}
\scalebox{0.68}{
\begin{tabular}{|l|c|c|cccc|}
\hline
\multirow{2}{*}{\textbf{Method}} & \multicolumn{1}{l|}{\multirow{2}{*}{\textbf{\# Generated Personas}}} & \multicolumn{1}{l|}{\multirow{2}{*}{\textbf{Average \# Items}}} & \multicolumn{4}{c|}{\textbf{Average \# Items in Sections}}                                                                                                 \\ \cline{4-7} 
                                 & \multicolumn{1}{l|}{}                                             & \multicolumn{1}{l|}{}                                      & \multicolumn{1}{c|}{\textbf{Jobs.}} & \multicolumn{1}{c|}{\textbf{Pain Pnts.}} & \multicolumn{1}{c|}{\textbf{Skills}} & \textbf{Tools} \\ \hline
\baseline\                       & 3 (1 per role)                                                    & 34.00                                                      & \multicolumn{1}{c|}{9.67}          & \multicolumn{1}{c|}{10.00}              & \multicolumn{1}{c|}{6.67}           & 7.67           \\ \hline
\woloop\                         & 3 (1 per role)                                                    & 32.33                                                      & \multicolumn{1}{c|}{9.33}          & \multicolumn{1}{c|}{9.00}               & \multicolumn{1}{c|}{7.00}           & 7.00           \\ \hline
\woresources\                    & 3 (1 per role)                                                    & 34.00                                                      & \multicolumn{1}{c|}{8.67}          & \multicolumn{1}{c|}{10.33}              & \multicolumn{1}{c|}{7.33}           & 7.67           \\ \hline
\approachsc\                     & 3 (1 per role)                                                    & 40.67                                                      & \multicolumn{1}{c|}{9.67}          & \multicolumn{1}{c|}{10.33}              & \multicolumn{1}{c|}{10.67}          & 10.00          \\ \hline
\end{tabular}
}
\vspace*{-.5em}
\end{table}
 
\subsubsection{RQ1 Results} 
Table~\ref{tab:rq1-persona-summary} summarizes the generated personas after expert validation. For each method, the table reports the total number of personas, the average number of items per persona, and the average number of items in each persona section. We analyze expert revisions in the interaction logs from two complementary perspectives: (1) review actions and (2) edits to the generated persona items.

\textbf{\emph{Review Action Analysis}.} Table~\ref{table:rq2_logs} reports the percentage of each action (edit, removal, addition, and approval) for \approachsc\ and the baselines. \approachsc\ achieves the highest approval rate (96.90\%), followed by \woloop\ (93.89\%) and \woresources\ (90.97\%). \baseline, which lacks both the critique-refinement loop and external resources, has the lowest approval rate (75.83\%).  Additions and removals occur only in \baseline; \woloop\ and both agentic approaches show 0\% additions and removals across sections. Among the four tools, \approachsc\ produces the fewest edits. Except for \baseline, none of the persona-generation methods produced a completely invalid item requiring removal.

\begin{table}[t]
\begin{center}
{\footnotesize
\caption{Percentage of persona items generated by \approach\ and the baselines that were edited, removed, or added.}
\label{table:rq2_logs}
\scalebox{1.}{
\begin{tabular}{l r r r r}
\toprule
\textbf{Method} & \textbf{Edit} & \textbf{Remove} & \textbf{Add} & \textbf{Approval} \\
\midrule
\baseline     & 9.12\% & 3.01\% & 12.04\% & 75.83\% \\
\woloop       & 6.11\% & 0\%    & 0\%     & 93.89\% \\
\woresources  & 9.03\% & 0\%    & 0\%     & 90.97\% \\
\approachsc  & 3.10\% & 0\%    & 0\%     & 96.90\% \\
\bottomrule
\end{tabular}
}
}
\end{center}
\vspace*{-2em}
\end{table}

\begin{figure}[t]
    \vspace*{-0.9em}
   \centering
   \includegraphics[width=0.95\linewidth]{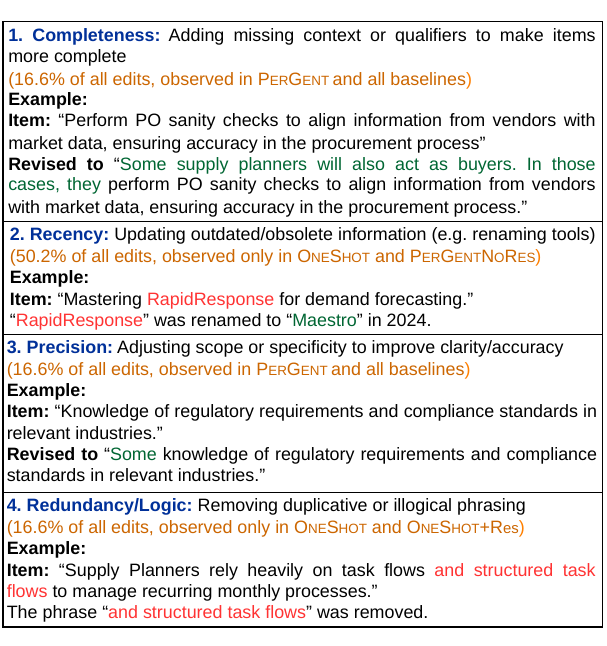}
   \vspace*{-0.3cm}
   \caption{Edit categories for generated persona items.}
   \label{fig:rq1_qualitative}
   \vspace*{-0.7cm}
\end{figure}

\textbf{\emph{Edit Content Analysis}.} We manually inspected experts' edits to the generated personas and categorized them into four types: \emph{Completeness}, \emph{Recency}, \emph{Precision}, and \emph{Redundancy/Logic}. Figure~\ref{fig:rq1_qualitative} presents these categories, including their definitions, frequencies, and examples. Recency-related edits account for nearly half of all modifications, making them the most common. Notably, such edits appear only in \baseline\ and \woresources. This pattern is expected: pretrained LLMs are trained on data collected only up to a fixed point in time (the knowledge cutoff)~\cite{cheng_2024_dated}. Without external resources, an LLM cannot reliably incorporate events or developments that occur after this cutoff. In contrast, \approachsc\ and \woloop\ prompted no recency edits because they can consult up-to-date external resources.

The remaining categories (completeness, precision, and redundancy/logic) each account for roughly 16\% of edits. These changes are comparatively minor and largely reflect contextual refinement rather than substantive issues in the personas. Completeness and precision edits often softened claims or added role-specific qualifiers, indicating that the personas were generally directionally correct but occasionally overgeneralized. Redundancy/logic edits primarily addressed style and internal coherence rather than conceptual gaps. Overall, aside from recency-related updates, the personas required limited semantic revision, suggesting strong alignment between the generated content and expert expectations.

\begin{tcolorbox}[breakable, colback=gray!10!white, colframe=black!75!black,
                  title=RQ1 Answer]
Across the generated personas, \approachsc\ achieved the highest expert approval rate (96.9\%). No items were removed or added when using \approachsc, and only 3.1\% required edits. Among the baselines, \baseline\ had the lowest approval rate (75.83\%), underscoring the negative impact of the simultaneous absence of both external resources and a critique--refinement loop. The remaining baselines, \woloop\ and \woresources, performed better than \baseline\ but still fell short of \approachsc. In general, we observed that most expert revisions addressed outdated information in approaches without external resource access.
\end{tcolorbox}

\subsection{RQ2 Experiments and Results} 
\label{subsec:RQ2}


\subsubsection{RQ2 Experiments} 
Figure~\ref{fig:exp_workflow} (bottom) shows the  workflow of the RQ2 experiments.
For RQ2, we generate personas using \approachsc\ and the baselines for the same job roles as the ten pre-LLM personas described in Section~\ref{subset:prellm}.
To generate each persona, we evaluate PerGent and each baseline with a single LLM at a time, selecting from seven LLMs: GPT-4o, o4-mini, \hbox{GPT-5.2}, Gemini-2, Gemini-3, Claude-4, or Grok-3.
To mitigate random variation in LLM outputs, we repeat each generation \hbox{\emph{five} times.} In total, our experiments produce 1400 personas (10 job roles $\times$ 4 methods $\times$ 7 LLMs $\times$ 5 repetitions). We compute the distinctness and preservation metrics defined in Section~\ref{subsec:metrics} by comparing each LLM-generated persona with its corresponding pre-LLM persona.


Due to the large number of comparisons required, manually measuring distinctness and preservation for all 1400 generated personas is impractical. We therefore employ an LLM -- commonly referred to as \emph{LLM-as-Judge}~\cite{Zheng_2023_LLM_as_Judge} -- to automatically perform these evaluations, following established best practices for using LLMs as automated evaluators~\cite{Huang_2025_iKnow, Parham_2025_Critique, Zheng_2023_LLM_as_Judge}. Our LLM-as-Judge is implemented using Gemini-2.

We recall from Section~\ref{subsec:metrics} that computing distinctness and preservation hinges on item-persona relationship classification. The LLM-as-Judge performs this classification and outputs both a label (``fully present'', ``partially present'', or ``not present'') and a brief rationale for each item-persona pair. We use these rationales to make the classifications auditable and to quantify alignment with Kinaxis experts, as discussed next.

Before deployment, we ensure that the LLM-as-Judge is a faithful proxy for human experts by calibrating it through iterative prompt refinement and validation. The full procedure is described in the supplementary material~\cite{Github}. After calibration, the LLM-as-Judge achieves an average Krippendorff's $\alpha$ of 0.865, indicating strong agreement with human experts~\cite{krippendorff_2018_content}. Experts also judged, on average, 96.5\% of the generated rationales as \emph{justified}. These results provide confidence that our LLM-as-Judge is a reliable proxy for expert assessment.



\subsubsection{RQ2 Results} Table~\ref{table:sec_cov} shows the distinctness and preservation scores between the personas generated by \approachsc\ and the baselines against their corresponding pre-LLM personas of the same job roles.  As seen from this table, personas generated by \approachsc\ have the highest average distinctness and full preservation. 
\begin{wrapfigure}{r}{0.39\columnwidth}
    \centering
    \vspace{-5pt} 
    \includegraphics[width=0.39\columnwidth]{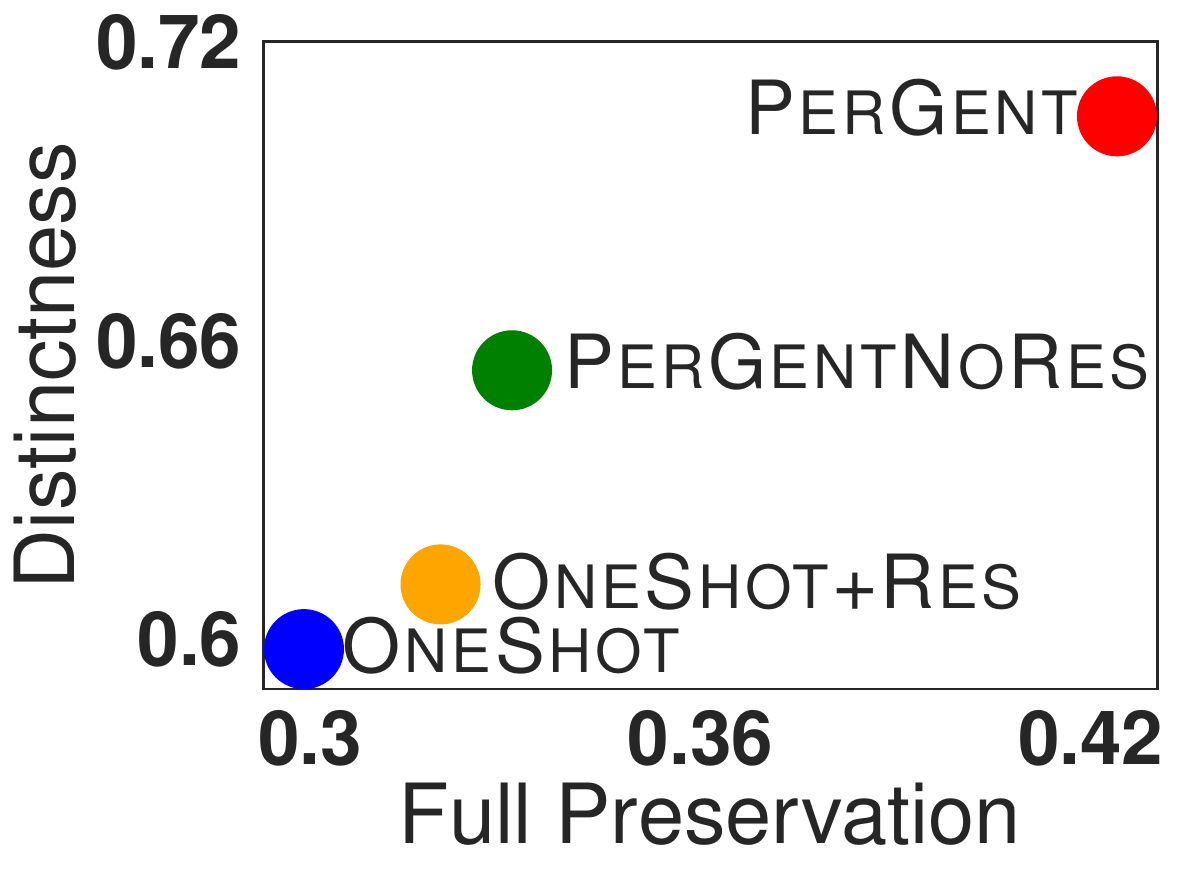}
    \caption{Distinctness vs. full preservation}
    \label{fig:rq2_plot}
\end{wrapfigure} 
Relative to \baseline, \approachsc\ improves distinctness and full preservation by 10.7\% and 12.5\%, respectively; the corresponding gains over \woloop\ are 9.4\% and 12.4\%, and over \woresources\ are 5.1\% and 9.3\%.
Figure~\ref{fig:rq2_plot} further illustrates the relationship between the average distinctness and full preservation for \approachsc\ and the baselines.


\begin{table}[!t]
\begin{center}
{\footnotesize
\caption{Distinctness and preservation results for different methods across different LLMs (averaged over five runs).}
\label{table:sec_cov}
\scalebox{0.75}{
\begin{tabular}{|
>{\centering\arraybackslash}p{2.1cm}|
>{\centering\arraybackslash}p{1.3cm}|
>{\centering\arraybackslash}p{1.3cm}|
>{\centering\arraybackslash}p{1.7cm}|
>{\centering\arraybackslash}p{1.7cm}|}
\hline
\textbf{Method} & \textbf{LLM} & \textbf{Distinctness} & \textbf{Full Preservation} & \textbf{Partial Preservation} \\ \hline

                                 & Claude-4  & 0.474 & 0.185 & 0.676 \\ \cline{2-5}
                                 & Gemini-2  & 0.526 & 0.305 & 0.613 \\ \cline{2-5}
                                 & Gemini-3  & 0.784 & 0.215 & 0.672 \\ \cline{2-5}
                                 & GPT-4o    & 0.529 & 0.424 & 0.574 \\ \cline{2-5}
                                 & Grok-3    & 0.485 & 0.304 & 0.613 \\ \cline{2-5}
                                 & o4-mini   & 0.589 & 0.345 & 0.646 \\ \cline{2-5}
                                 & GPT-5.2   & 0.794 & 0.310 & 0.685 \\ \cline{2-5}
\multirow{-8}{*}{\textbf{\baseline}} 
                                 & \cellcolor[HTML]{D0D0D0}Average 
                                 & \cellcolor[HTML]{D0D0D0}0.598 
                                 & \cellcolor[HTML]{D0D0D0}0.299 
                                 & \cellcolor[HTML]{D0D0D0}0.640 \\ \hline

                                 & Claude-4                      & 0.475         & 0.317 & 0.683 \\ \cline{2-5}
                                 & Gemini-2                      & 0.581         & 0.303 & 0.697 \\ \cline{2-5}
                                 & Gemini-3                      & 0.848         & 0.321 & 0.705 \\ \cline{2-5}
                                 & GPT-4o                        & 0.506         & 0.366 & 0.634 \\ \cline{2-5}
                                 & Grok-3                        & 0.466         & 0.306 & 0.694 \\ \cline{2-5}
                                 & o4-mini                       & 0.584         & 0.262 & 0.738 \\ \cline{2-5}
                                 & GPT-5.2                       & 0.813         & 0.366 & 0.602 \\ \cline{2-5}
\multirow{-8}{*}{\textbf{\woloop}} 
                                 & \cellcolor[HTML]{D0D0D0}Average 
                                 & \cellcolor[HTML]{D0D0D0}0.611 
                                 & \cellcolor[HTML]{D0D0D0}0.320 
                                 & \cellcolor[HTML]{D0D0D0}0.679 \\ \hline

                                 & Claude-4                      & 0.676         & 0.359 & 0.641 \\ \cline{2-5}
                                 & Gemini-2                      & 0.439         & 0.265 & 0.735 \\ \cline{2-5}
                                 & Gemini-3                      & 0.927         & 0.150 & 0.463 \\ \cline{2-5}
                                 & GPT-4o                        & 0.496         & 0.346 & 0.654 \\ \cline{2-5}
                                 & Grok-3                        & 0.445         & 0.448 & 0.552 \\ \cline{2-5}
                                 & o4-mini                       & 0.698         & 0.478 & 0.522 \\ \cline{2-5}
                                 & GPT-5.2                       & 0.895         & 0.272 & 0.454 \\ \cline{2-5}
\multirow{-8}{*}{\textbf{\woresources}} 
                                 & \cellcolor[HTML]{D0D0D0}Average 
                                 & \cellcolor[HTML]{D0D0D0}0.654 
                                 & \cellcolor[HTML]{D0D0D0}0.331 
                                 & \cellcolor[HTML]{D0D0D0}0.574 \\ \hline

                                 & Claude-4                      & 0.815         & 0.388 & 0.712 \\ \cline{2-5}
                                 & Gemini-2                      & 0.522         & 0.239 & 0.761 \\ \cline{2-5}
                                 & Gemini-3                      & 0.933         & 0.254 & 0.705 \\ \cline{2-5}
                                 & GPT-4o                        & 0.510         & 0.433 & 0.667 \\ \cline{2-5}
                                 & Grok-3                        & 0.626         & 0.494 & 0.606 \\ \cline{2-5}
                                 & o4-mini                       & 0.614         & 0.496 & 0.604 \\ \cline{2-5}
                                 & GPT-5.2                       & 0.917         & 0.435 & 0.691 \\ \cline{2-5}
\multirow{-8}{*}{\textbf{\approachsc}} 
                                 & \cellcolor[HTML]{D0D0D0}Average 
                                 & \cellcolor[HTML]{D0D0D0}0.705 
                                 & \cellcolor[HTML]{D0D0D0}0.424 
                                 & \cellcolor[HTML]{D0D0D0}0.678 \\ \hline

\end{tabular}

}}
\end{center}
\vspace*{-.4cm}
\end{table}

\begin{table}[t]
\begin{center}
{\footnotesize
\caption{Statistical tests comparing \approachsc\ to baselines; $p$-values evaluated at 95\% confidence level.}
\label{tab:stats}
\scalebox{.72}{
\begin{tabular}{|c|
>{\centering\arraybackslash}p{.9cm}
>{\centering\arraybackslash}p{.95cm}|
>{\centering\arraybackslash}p{.9cm}
>{\centering\arraybackslash}p{1.1cm}|
>{\centering\arraybackslash}p{.9cm}
>{\centering\arraybackslash}p{1.1cm}|}
\hline
\multirow{2}{*}{\textbf{\scriptsize Comparing \approachsc\ with:}} 
& \multicolumn{2}{c|}{\textbf{Distinctness}} 
& \multicolumn{2}{c|}{\textbf{Full Preservation}} 
& \multicolumn{2}{c|}{\textbf{Partial Preservation}} \\ \cline{2-7} 

& \multicolumn{1}{c|}{\textbf{$p$-value}} & \textbf{$\hat{A}_{12}$}
& \multicolumn{1}{c|}{\textbf{$p$-value}} & \textbf{$\hat{A}_{12}$}
& \multicolumn{1}{c|}{\textbf{$p$-value}} & \textbf{$\hat{A}_{12}$} \\ 
\hline

\textbf{\baseline} 
& \multicolumn{1}{c|}{\textbf{0.000}} & \cellcolor{green!20} 0.73 (L)
& \multicolumn{1}{c|}{\textbf{0.000 }} & \cellcolor{green!20} 0.72 (L)
& \multicolumn{1}{c|}{\textbf{0.000}} & \cellcolor{cyan!20} 0.67 (M) \\ 
\hline

\textbf{\woloop} 
& \multicolumn{1}{c|}{\textbf{0.000 }} & \cellcolor{green!20} 0.72 (L)
& \multicolumn{1}{c|}{\textbf{0.000 }} & \cellcolor{green!20} 0.72 (L)
& \multicolumn{1}{c|}{0.976} & 0.50 (N) \\ 
\hline

\textbf{\woresources} 
& \multicolumn{1}{c|}{\textbf{0.000 }} & \cellcolor{yellow!40} 0.60 (S)
& \multicolumn{1}{c|}{\textbf{0.000 }} & \cellcolor{cyan!20} 0.70 (M)
& \multicolumn{1}{c|}{\textbf{0.000}} & \cellcolor{green!20} 0.76 (L) \\ 
\hline

\end{tabular}
}
}
\end{center}
\flushleft
\footnotesize{
$^\ast$ Effect sizes: Large \colorbox{green!20}{\makebox(1,1){}}, Medium \colorbox{cyan!20}{\makebox(1,1){}}, Small \colorbox{yellow!40}{\makebox(1,1){}}. $p \leq 0.05$ in bold.
}
\vspace*{-.4cm}
\end{table}

Table~\ref{tab:stats} reports the Wilcoxon signed-rank test~\cite{wilcoxon} $p$-values (statistical significance) alongside the Vargha–Delaney $\hat{A}_{12}$ effect size~\cite{vargha} for pairwise comparisons of \approachsc\ against each baseline on the distinctness and preservation metrics. Following Vargha and Delaney~\cite{vargha}, we classify effect sizes into four levels:
Negligible ($0.44 \le \hat{A}_{12} < 0.56$),
Small ($0.56 < \hat{A}_{12} \le 0.64$ or $0.36 \le \hat{A}_{12} < 0.44$),
Medium ($0.64 < \hat{A}_{12} \le 0.71$ or $0.29 \le \hat{A}_{12} < 0.36$),
and Large ($\hat{A}_{12} > 0.71$ or $\hat{A}_{12} < 0.29$).

We observe from Table~\ref{tab:stats} that all comparisons between \approachsc\ and its baselines are statistically significant at the 95\% confidence level, except for partial preservation against \woloop\ ($p=0.976$). 
\approachsc\ significantly outperforms the baselines on distinctness and full preservation with medium to large effect sizes in five of the six comparisons. 
It also significantly outperforms \woresources\ on distinctness, though with a small effect size.
For partial preservation, \approachsc\ significantly outperforms \baseline\ and \woresources\ by 3.8\% and 10.4\%, respectively, and is virtually tied with \woloop\ (a 0.1\% advantage for \woloop), which, as noted earlier, is the only statistically insignificant comparison.

Finally, Table~\ref{table:sec_cov} shows substantial variation across underlying LLMs for every approach. For instance, under \approachsc, distinctness ranges from 51.0\% to 93.3\%, and full preservation ranges from 23.9\% to 49.6\% across LLMs. This pattern indicates that, regardless of the persona-generation approach, the choice of the underlying LLM strongly influences the quality of the generated personas.


\begin{tcolorbox}[breakable, colback=gray!10!white, colframe=black!75!black,
                  title=RQ2 Answer]
Relative to the pre-LLM personas, \approachsc\ is able to \emph{reproduce a meaningful proportion of expert content while also contributing substantial new content}. Averaged over the ten persona roles (7 LLMs, 5 runs), \approachsc\ achieves $\approx$0.71 \emph{distinctness} -- i.e., much of what it generates is not already covered in the expert personas -- while fully preserving $\approx$0.42 of expert items and partially preserving $\approx$0.68 (Table~\ref{table:sec_cov}). Compared to the baselines, \approachsc\ ranks the highest on \emph{distinctness} and \emph{full preservation} among all methods, while its \emph{partial preservation} is on par with the strongest baseline (\woloop). We also observe \emph{considerable variation across LLMs}, indicating that model choice strongly affects outcomes. Finally, preservation remains well below 1.0, indicating that \emph{pre-LLM personas contain a non-negligible amount of expert information not covered by LLM generation}; we return to this gap in \emph{Lessons Learned}.
\end{tcolorbox}

\subsection{RQ3 Experiments and Results} 
\label{subsec:RQ3}
\subsubsection{RQ3 Experiments}
Figure~\ref{fig:exp_workflow} (bottom) shows the workflow of the RQ3 experiments.
To answer RQ3, within the RQ2 experimental setup, we measure for \approachsc\ and the baselines: (i) the number of LLM calls made to generate a persona and (ii) the number of input and output tokens consumed per call.

\subsubsection{RQ3 Results} 
Table~\ref{table:costs} reports, for each method, the average number of LLM calls and the average input and output tokens per call (computed over ten roles, seven LLMs, and five repetitions).
As the table shows, across all models, \baseline\ uses only a single compact call per persona (315 input/353 output tokens on average). \woloop\ also uses one call but, by adding external resources without critique-refinement, raises the average input size by about an order of magnitude (3,198.9 tokens) while increasing output size only modestly (390 tokens), indicating that resources mainly increase prompt length rather than response verbosity. In contrast, \woresources\ and \approachsc\ use multiple generator-critic interactions, averaging 5.1 and 4.8 LLM calls per persona, respectively. This orchestration substantially increases cumulative token use: input tokens grow because each critique-refinement iteration includes the full instructions and accumulated history, and, for \approachsc, external resources (see Section~\ref{sec:methodology}); output tokens grow because the methods produce intermediate drafts and critiques \hbox{in addition to the final persona.}


\begin{table}[t]
\begin{center}
{\footnotesize
\caption{\textit{Cost} results. The average number of LLM calls and average number of tokens in the input prompt and output returned in each call for \approachsc\ and its variants.}
\label{table:costs}
\scalebox{0.81}{
\begin{tabular}{|c|c|cc|c|}
\hline
                                    & \multicolumn{1}{c|}{}                               & \multicolumn{2}{c|}{\textbf{Average \# Tokens}}                                        & \multicolumn{1}{c|}{}                                             \\ \cline{3-4}
\multirow{-2}{*}{\textbf{Method}}  & \multicolumn{1}{c|}{\multirow{-2}{*}{\textbf{LLM}}} & \multicolumn{1}{c|}{\textbf{Input}}                  & \textbf{Output}                 & \multicolumn{1}{c|}{\multirow{-2}{*}{\textbf{Avg. \# Calls}}} \\ \hline
                                    & Claude-4                                            & \multicolumn{1}{c|}{308}                             & 194.1                           & 1                                                                 \\ \cline{2-5} 
                                    & Gemini-2                                            & \multicolumn{1}{c|}{328}                             & 266.3                           & 1                                                                 \\ \cline{2-5} 
                                    & Gemini-3                                            & \multicolumn{1}{c|}{328}                             & 470.0                           & 1                                                                 \\ \cline{2-5} 
                                    & GPT-4o                                              & \multicolumn{1}{c|}{306}                             & 270.8                           & 1                                                                 \\ \cline{2-5} 
                                    & GPT-5.2                                             & \multicolumn{1}{c|}{306}                             & 491.7                           & 1                                                                 \\ \cline{2-5} 
                                    & Grok-3                                              & \multicolumn{1}{c|}{323}                             & 315.3                           & 1                                                                 \\ \cline{2-5} 
                                    & o4-mini                                             & \multicolumn{1}{c|}{306}                             & 463.1                           & 1                                                                 \\ \cline{2-5} 
\multirow{-8}{*}{\textbf{\baseline}} & \cellcolor[HTML]{C0C0C0}Average                     & \multicolumn{1}{c|}{\cellcolor[HTML]{C0C0C0}315.0}   & \cellcolor[HTML]{C0C0C0}353.0  & \cellcolor[HTML]{C0C0C0}1                                         \\ \hline
                                    & Claude-4                                            & \multicolumn{1}{c|}{3205}                            & 189.9                           & 1                                                                 \\ \cline{2-5} 
                                    & Gemini-2                                            & \multicolumn{1}{c|}{3239}                            & 305.2                           & 1                                                                 \\ \cline{2-5} 
                                    & Gemini-3                                            & \multicolumn{1}{c|}{3239}                            & 519.2                           & 1                                                                 \\ \cline{2-5} 
                                    & GPT-4o                                              & \multicolumn{1}{c|}{3160}                            & 291.7                           & 1                                                                 \\ \cline{2-5} 
                                    & GPT-5.2                                             & \multicolumn{1}{c|}{3160}                            & 586.9                           & 1                                                                 \\ \cline{2-5} 
                                    & Grok-3                                              & \multicolumn{1}{c|}{3229}                            & 321.9                           & 1                                                                 \\ \cline{2-5} 
                                    & o4-mini                                             & \multicolumn{1}{c|}{3160}                            & 515.1                           & 1                                                                 \\ \cline{2-5} 
\multirow{-8}{*}{\textbf{\woloop}}   & \cellcolor[HTML]{C0C0C0}Average                     & \multicolumn{1}{c|}{\cellcolor[HTML]{C0C0C0}3198.9}  & \cellcolor[HTML]{C0C0C0}390.0  & \cellcolor[HTML]{C0C0C0}1                                         \\ \hline
                                    & Claude-4                                            & \multicolumn{1}{c|}{3792.9}                          & 1656                            & 5.7                                                               \\ \cline{2-5} 
                                    & Gemini-2                                            & \multicolumn{1}{c|}{5911.8}                          & 1004.3                          & 5.4                                                               \\ \cline{2-5} 
                                    & Gemini-3                                            & \multicolumn{1}{c|}{8629.2}                          & 1739.1                          & 5.0                                                               \\ \cline{2-5} 
                                    & GPT-4o                                              & \multicolumn{1}{c|}{5694.6}                          & 976.1                           & 3.5                                                               \\ \cline{2-5} 
                                    & GPT-5.2                                             & \multicolumn{1}{c|}{13318.1}                         & 2205.6                          & 4.47                                                              \\ \cline{2-5} 
                                    & Grok-3                                              & \multicolumn{1}{c|}{8919.6}                          & 1656.3                          & 6.1                                                               \\ \cline{2-5} 
                                    & o4-mini                                             & \multicolumn{1}{c|}{11227.1}                         & 2018.9                          & 5.5                                                               \\ \cline{2-5} 
\multirow{-8}{*}{\textbf{\woresources}} & \cellcolor[HTML]{C0C0C0}Average                   & \multicolumn{1}{c|}{\cellcolor[HTML]{C0C0C0}8213.3}  & \cellcolor[HTML]{C0C0C0}1608.0 & \cellcolor[HTML]{C0C0C0}5.1                                       \\ \hline
                                    & Claude-4                                            & \multicolumn{1}{c|}{6650.5}                          & 645.9                           & 4.3                                                               \\ \cline{2-5} 
                                    & Gemini-2                                            & \multicolumn{1}{c|}{9105.2}                          & 1249.5                          & 4.9                                                               \\ \cline{2-5} 
                                    & Gemini-3                                            & \multicolumn{1}{c|}{13064.2}                         & 1804.7                          & 4.2                                                               \\ \cline{2-5} 
                                    & GPT-4o                                              & \multicolumn{1}{c|}{8581.7}                          & 993.7                           & 3.7                                                               \\ \cline{2-5} 
                                    & GPT-5.2                                             & \multicolumn{1}{c|}{14601.5}                         & 2494.9                          & 4.4                                                               \\ \cline{2-5} 
                                    & Grok-3                                              & \multicolumn{1}{c|}{11091.4}                         & 2161.7                          & 5.8                                                               \\ \cline{2-5} 
                                    & o4-mini                                             & \multicolumn{1}{c|}{13506.5}                         & 2205.6                          & 6.2                                                               \\ \cline{2-5} 
\multirow{-8}{*}{\textbf{\approachsc}} & \cellcolor[HTML]{C0C0C0}Average                    & \multicolumn{1}{c|}{\cellcolor[HTML]{C0C0C0}10943.0} & \cellcolor[HTML]{C0C0C0}1650.9 & \cellcolor[HTML]{C0C0C0}4.8                                       \\ \hline
\end{tabular}

}
}
\vspace*{-.6cm}
\end{center}
\end{table}

\begin{tcolorbox}[breakable, colback=gray!10!white, colframe=black!75!black,
                  title=RQ3 Answer]
On average, \approachsc\ requires \emph{4.8 LLM calls per persona}, with 11K input tokens and 1.7K output tokens \emph{per call} (Table~\ref{table:costs}). In contrast, the single-shot baselines use \emph{one} call: \baseline\ averages 315/353 input/output tokens, and \woloop\ averages 3.2K/390. The agentic variant without external resources (\woresources) has a similar number of calls (5.1) but fewer input tokens (8.2K).
\end{tcolorbox}

\subsection{Validity Considerations}\label{sec:threats}


\textbf{Internal validity.} Expert-specific factors (e.g., editing style, strictness, domain focus) and learning effects could confound comparisons; we mitigate such confounding by recruiting experts with comparable seniority and using a balanced cross-over design. Measurement bias could arise if persona changes are missed; we mitigate this by logging every generation and every update/approval during the one-month in-situ study. Nevertheless, individual reviewing behaviours may still affect the observed edit and approval rates, given the limited number of expert interactions.


\textbf{Construct validity.} Update rates and approvals are proxies for usefulness and post-edit effort but may not fully capture correctness; we mitigate this threat by logging all item-level actions and manually inspecting and categorizing edits to confirm they address quality issues. Preservation and distinctness are defined relative to expert-created pre-LLM personas, which may reflect expert assumptions, organizational conventions, omissions, temporal drift, or gaps/skews from limited direct input from practitioners; we mitigate this threat to the extent possible via expert review and removal of outdated/irrelevant pre-LLM items. LLM-as-Judge is a proxy for expert judgment; we mitigate potential misalignment via iterative prompt calibration and validation against experts and observing strong agreement (Krippendorff's $\alpha=0.865$). Still, the judge should be viewed as an approximation of expert assessment rather than a replacement for it.

\textbf{Conclusion validity.} Our expert study is small and yielded few edits. RQ1 should therefore be interpreted as practice-based evidence from a one-month deployment rather than as a basis for strong claims. The observed differences may not replicate with other experts or roles; we mitigate this with a balanced cross-over design, but the low number of editing events limits statistical power and makes our analysis sensitive to noise. Although RQ2 evaluates 1,400 LLM-generated personas, its statistical results remain tied to the studied reference roles and should not be read as supporting broader claims.

\textbf{External validity.} Our results are based on one organization, domain, and persona format, and may not generalize to other industries, persona structures, or settings with different resources. The study involved 11 experts at Kinaxis, as outlined in Sections~\ref{sec:context}, \ref{subsec:rq1}, and \ref{subsec:RQ2}. RQ1 involved 12 expert-reviewed personas. RQ2 used 10 pre-LLM personas as references for evaluating 1,400 LLM-generated personas. The number of reference roles is thus limited despite the larger number of evaluated personas. While our evaluation provides promising evidence from a single case study, further case studies across other organizations, domains, persona structures, and systems remain essential to establish the broader applicability and usefulness of our findings.

\subsection{Data Availability}\label{sec:availability}
All our publicly shareable artifacts, including the code for \approachsc\ and its baselines (\baseline, \woloop, and \woresources), prompt outlines, the code for our experimental procedures, evaluation results, additional expert-edited examples, sample personas generated by \approachsc, and the detailed process for calibrating the LLM-as-Judge with experts, are available online~\cite{Github}. Note that the prompts and code in our online repository correspond to the public-data version of the implementation and exclude Kinaxis-specific customizations that cannot be shared due to confidentiality.

\section{Related Work}
\label{sec:relwork}

Traditional persona generation is largely manual, relying on designers and domain experts to synthesize qualitative insights into user archetypes~\cite{Cooper1999Inmates}. Early automation attempts were mostly limited to data-driven techniques that infer personas from behavioural traces such as social-media analytics and interaction logs~\cite{An_2018_Imaginary_People,Jansen_2020_Personas_and_analytics}, but these approaches are constrained by their reliance on predominantly quantitative data. In practice, however, personas must also capture workflow descriptions, task requirements, and cross-functional responsibilities, which are typically expressed in textual data. Therefore, methods that can process and synthesize text are needed, motivating the use of LLMs for persona construction.

Recently, several approaches have used LLMs to generate personas, but they largely rely on single-shot prompting without a critique-refinement loop. Jung et al.~\cite{Salminen_2025_PersonaCraft} use one-pass LLM generation to convert clustered survey responses into persona narratives, while Zhang et al.~\cite{Chetan_2023_PersonaGen} apply a similar single-step approach to extract persona traits from requirements-related feedback. Because LLMs can introduce plausible-sounding but ungrounded details, evaluation and validation are central to LLM-based persona generation. Existing evaluation of LLM-generated personas is largely perception-based: Salminen et al.~\cite{salminen2024deus} evaluate personas with internal evaluators and subject-matter experts using human-rated persona-quality dimensions, while Schuller et al.~\cite{schuller2024llm} compare LLM- and human-written personas in a human-subject study and find similar perceived quality and acceptance, with participants unable to reliably distinguish the personas' origin. However, these perception-based evaluations provide limited evidence of real-world usefulness. Our work addresses two major gaps in prior work on LLM-based persona generation. First, we introduce an agentic critique-refinement process that iteratively evaluates and improves generated personas. Second, we examine the practical usefulness of generated personas through two forms of industry evidence: (1) an in-situ deployment at Kinaxis to assess, during day-to-day use, how well generated personas meet expert expectations; and (2) a comparison with pre-LLM personas developed using industry best practices.
\section{Lessons Learned and Conclusion}
\label{sec:lessons}
Our collaboration with Kinaxis and the deployment of \approach\ in an industrial context led to some noteworthy lessons for both researchers and practitioners.

\textbf{Lesson 1: The critique-refinement loop is indispensable when external resources are unavailable.} Across both evaluations (expert in-situ review and comparison to pre-LLM expert personas), the best results arise when external resources are combined with an agentic critique-refinement loop. Nevertheless, when up-to-date domain artifacts (e.g., interviews and surveys) are available, single-shot generation already performs reasonably well, and the critique-refinement loop yields a smaller, incremental improvement. \emph{In contrast}, when such resources are absent, single-pass generation deteriorates substantially. In this setting, the critique-refinement loop becomes key to quality.
The practical take-away here, based on our findings for RQ2 and RQ3, is:

\begin{itemize}
    \item Use external resources alongside a critique-refinement loop for best results; 
    however, when high-quality external resources exist, the improvement from the critique-refinement loop is comparatively modest. 
    These gains should be weighed against added inference cost: \approach\ averaged 4.8 LLM calls per persona, with longer prompts from critique-refinement loop and external resources, versus one LLM call for non-agentic baselines.
    \item If high-quality external resources are unavailable or expensive to curate, ensure the critique-refinement loop is used, as it compensates to a meaningful extent for the absence of external resources.
\end{itemize}

\textbf{Lesson 2: LLMs can be a double-edged sword for completeness.}
In our in-situ deployment (RQ1), the stronger methods produced drafts that experts largely approved, with no additions made by the experts. This suggests that reviewers perceived few (if any) major omissions during the review. Yet, the comparison with pre-LLM personas (RQ2) reveals a clear tension: preservation remains well below 1.0 (e.g., full preservation $\approx$\,0.42 for \approach), even though distinctness is high ($\approx$\,0.71). In other words, despite being content-rich, the LLM-generated personas fail to capture a significant portion of expert knowledge. A plausible explanation for what may have happened in the in-situ study is an \emph{interaction effect} introduced by the review setting itself. Because experts were presented with a rich draft rather than starting from a blank page, the task may have implicitly shifted from \emph{elicitation} to \emph{evaluation}. In such settings, reviewers may naturally focus on improving and validating what is already present (e.g., refining wording or qualifiers) rather than actively searching for missing information. Consequently, omissions that might have surfaced during a blank-page analysis may remain unnoticed during draft review. While our data does not isolate the underlying root cause, it suggests an important implication for requirements completeness: \emph{expert approval of an LLM-generated artifact should not be interpreted as evidence of completeness.} In practice, when LLM outputs serve as the starting point for analysis, completeness may need to be assessed as an explicit objective rather than inferred from expert acceptance alone.

\textbf{Lesson 3: The LLM you choose matters.}
Despite increasing convergence on standard benchmarks, the evaluated LLMs showed substantial task-specific variation in our setting. For example, under \approachsc, distinctness ranged from 0.51 to 0.93, and full preservation ranged from 0.24 to 0.50 across LLMs (Table~\ref{table:sec_cov}). This indicates that LLMs are not interchangeable in practice: the same method can yield materially different outcomes depending on the model and the evaluation criteria. The key takeaway is that LLM selection remains an empirical decision requiring task-specific evaluation in the target context using appropriate data and metrics, rather than extrapolation from benchmark rankings.

\section*{Acknowledgements}
We gratefully acknowledge financial support from Mitacs, Kinaxis, and NSERC of Canada through the Discovery Program. We thank Gelu Ticala for his support and feedback.


\begingroup
\balance
\bibliographystyle{IEEEtran}
\bibliography{references}
\endgroup

\clearpage
\end{document}